
\newskip\oneline \oneline=1em plus.3em minus.3em
\newskip\halfline \halfline=.5em plus .15em minus.15em
\newbox\sect
\newcount\eq
\newbox\lett

\def\simlt{\mathrel{\lower2.5pt\vbox{\lineskip=0pt\baselineskip=0pt
           \hbox{$<$}\hbox{$\sim$}}}}
\def\simgt{\mathrel{\lower2.5pt\vbox{\lineskip=0pt\baselineskip=0pt
           \hbox{$>$}\hbox{$\sim$}}}}

\newdimen\short
\def\adv{\global\advance\eq by1}
\def\set#1#2{\setbox#1=\hbox{#2}}
\def\nextlet#1{\global\advance\eq by-1\setbox
                \lett=\hbox{\rlap#1\phantom{a}}}

\newcount\eqncount
\eqncount=0
\def\equn{\global\advance\eqncount by1\eqno{(\the\eqncount)} }
\def\put#1{\global\edef#1{(\the\eqncount)}           }

\def\adler{1}
\def\weinbf{2}
\def\weinbs{3}
\def\addashen{4}
\def\schnitzer{5}
\def\bels{6}
\def\tntf{7}
\def\im{8}
\def\tnts{9}
\def\callan{10}
\def\bel{11}
\def\omnes{12}
\def\ksrf {13}
\def\meisner{14}

\def\marshak{15}
\def\gasser{16}
\def\estab{17}
\def\exps{18}
\def\pdg{19}
\def\expf{20}
\def\donald{21}

\magnification=1200
\hsize=6.0 truein
\vsize=8.5 truein
\baselineskip 14.5pt

\overfullrule=0 pt
\nopagenumbers
\rightline{hep-ph/9411423}
\rightline{CPTH-A294.0294}
\rightline{November 1994}
\vskip 1.0truecm

\centerline{\bf  $\tau \to \pi K \nu$ DECAY AND $\pi K$ SCATTERING}
\vskip 1.0truecm
\centerline{{\bf L. Beldjoudi} and {\bf Tran N. Truong}}
\vskip .5truecm

\centerline{{\it Centre de Physique Th{\'e}orique}
\footnote{$^*$}{\it Laboratoire Propre du CNRS UPR A.0014}}
\centerline{\it Ecole Polytechnique, 91128 Palaiseau, France}
\vskip 2.5truecm
\centerline{\bf ABSTRACT}
\vskip .5truecm

Using  chiral low energy theorems and  elastic unitarity assumption,
 the $\tau\to\pi K \nu $ decay is investigated. The vector and scalar
$\pi K$ form factors are calculated.
It is found that the $\pi K$ spectrum is dominated by the $K^*$ resonance.
By measuring the forward-backward asymmetry, it is shown that the
S wave $\pi K$ phase shift can be determined near the $K^{*}$ resonance region.
The calculated branching ratio and  resonance parameters are in good agreement
with experiments.

\vskip 3.5truecm

\vfill\eject

\footline={\hss\tenrm\folio\hss}\pageno=1

\vskip 1.0 cm

\centerline{\bf {INTRODUCTION}}
\vskip 0.5 cm

Current Algebra  was invented in the 60's to study phenomena involving
emission of
soft pions (and kaons) and also  the chiral symmetry breaking effects.
Its first success was the study of the renormalization effect of the axial
vector nucleon coupling constant $g_A$ due to the strong interaction by the
celebrated Adler-Weisberger relation which relates  $ g_A$ to the $\pi  N$
total cross sections[\adler].  Subsequent calculations by Weinberg
[\weinbf, \weinbs]
 and others [\addashen] on
the soft pion phenomena,  such as the pion nucleon scattering lengths the
relation between $K_{l2},K_{l3}$ and  $K_{l4}$ etc.
confirmed the success of current algebra as an useful tool to study the low
energy pion physics. It was later realized that the pions emitted in these
processes are not really soft and methods were invented to take into account of
the correction to the soft pion current  algebra theorems.
This development was known as the hard pion current algebra which consists in
supplementing the low energy current algebra theorems with unitarity
corrections in the
form of the pole dominance for the hadronic matrix elements.

One of the ambitious program  vigorously pursued in the late 60's  was the
$\pi \rho A_1 $ system [\schnitzer] which ended up in failure due to the
wrong prediction of
the $A_1$ width. We shall deal with this problem in a future publication
[\bels]; we
discuss in this article a  simpler problem  $\tau \to \pi K \nu$ decay.
Unlike in the study of the pion electromagnetic form factor, where the relevant
current is exactly conserved  and where chiral symmetry does not play a role in
deriving the low energy theorems,
we deal here with the $ \pi K$ vector form factors whose current is not
exactly conserved due to the approximate SU(3) symmetry; the physics
is therefore richer.
Chiral symmetry does play an important role here which enables us to derive
the SU(3) breaking relation of the ratio $ f_K/f_\pi$ in terms of the two form
factors of the $K_{l3}$ decay [\callan].
These form factors are, on general grounds,  analytic  in the
momentum transfer s plane with a cut from $(m_\pi+m_K)^2$ to infinity. The
measured
 $ K_{l3}$ form factors give us only information on the form factors below
the cut.
In contrast,  the $
\tau \to \pi K \nu$ decay form factors are measured on the cut. They are
therefore the
analytic continuation of the $K_{l3}$ form factors to the time like region.
The role of the square
root threshold singularity in the scalar form factor was emphasized by one
of us
and provided a semi
quantitative understanding why the soft pion theorems are valid in some
reactions but not in
 others [\tntf].

The $\tau \to\pi K\nu$ decay amplitudes satisfy the same low energy
theorems as those of the
$Kl_{3}$ decay because they are the analytic continuation of each other.
The practical problem
is how to carry out the analytic continuation.

This problem was addressed along
time ago by using the boundary conditions of the low energy current algebra
theorems together with analyticity and elastic unitarity
relation[\marshak]. A singular
integral equation of the Muskhelishvili Omn\`{e}s (MO) type can be written
[\omnes] and whose
exact solution can be written in terms of the $I=1/2$ S and P wave
$ \pi K$ phase shifts.

One can either use experimental data or theoretical calculation of the $\pi K$
phase shifts in  the
solution of these integral equations in order to calculate the $\pi K$ form
factors.
This was done in the
reference [\tntf]  where both $K_{l3}$  form factors   were calculated.

An alternative method consists in writing an integral equation for the
inverse of
the form factor, similarly to the study of the pion vector and scalar form
factors [\tnts], we then get an approximate solution by solving it
perturbatively using
the $\pi K$ rms radius as input. The final solution satisfies the elastic
unitarity relation
and can take into account of the resonant
or non resonant interactions which were well demonstrated in the pion form
factor calculation. This result is equivalent to applying the Pad\'{e}
approximant method to the one loop chiral perturbation theory (CPTh) which
in the P wave
case leads to a $\rho$ resonance [\tnts].
The once iterated solution of the inverse integral equation or the Pad\'e
method
are in fact the bubble summation of
the $\pi\pi$ interaction of the form factor problem. This
approximation is now known in the litterature as the large $N_f$ method
[\im], where $N_f$ is
the number of Nambu-Goldstone bosons. It can be straightforwardly shown
that if the
strong partial wave amplitude could be represented by the bubble summation, the
Pad\'{e} method for the form factor would be the exact solution of the MO
integral equation.
 We want to emphasize that the exact
solution of the MO integral in terms of the phase shift is more general.

We show in this paper, using the  rms radii of the vector and scalar   $\pi K$
 form factors which are either given by  the experimental data
or by the Callan Treiman relation [\callan], the main
features of the $\tau \to \pi K \nu$ decay are completely determined.
  We wish to emphasize that the CPTh which was invented to study the
physics near
the $\pi K$ threshold cannot handle the main feature of
 the vector $
\pi K$ form factor because it cannot take into account of
the $K^{*}$ resonance. As we show below, our calculation for this decay
mode yields a correct $\pi K$ spectrum and a branching ratio of $1.0\%$
which is in agreement with the
experimental data of $1.4\pm 0.2 \%$.

We then improve the above calculations with a more accurate calculation of the
form factors where the t and u channels contributions to the $\pi K \to \pi K
$ amplitudes are taken into account as a correction.

This paper is organized as follows: in section 1 we give the
kinematic of the problem and a general phenomenological method to determine
the S wave phase shift by measuring the Forward-Backward asymmetry of the
$\pi K$ system.
We also include, for completeness, a short review of the analysis $K \to
\pi e \nu$
decay, together with the current algebra results for the form factors.

In section 2, the one loop  correction to current algebra result is
given and then  this result is modified to take into account of the elastic
unitarity condition in the approximation where the left hand cut
contribution to the $\pi K$
scattering amplitude is neglected.  In section 3, a more exact
calculation is presented where
the left hand cut contribution to the $\pi K$ scattering amplitude is taken
into account. A comparison between the two methods will be made.

\vskip 1.0 cm

{\bf I) Notations and kinematical preliminaries}

\vskip 0.5 cm

The most general $\tau \to \pi K \nu$ decay amplitudes are given in terms of
two form factors: $$\langle\pi ^{0} K^{-}\vert V_{\mu} ^{4-i5}(0)\vert0
\rangle=f_1(s)(p_2-p_1)_\mu+f_2(s)(p_1+p_2)_\mu  \kern 1.5 cm (1) $$
where $ p_1$, and $ p_2$ are, respectively,  the  pion and kaon momenta,
and $ s=(p_1+p_2)^2$ is
the time-like  momentum transfer  and  $V_{\mu}^{4-i5}$ is the vector
current operator with the
superscript indices referring to the SU(3) octet currents.
 $f_1(s)$ is the P wave $\pi K$ form factor,  $f_2(s)$ is a linear
combination of S and P states as can be
seen by taking the divergence of Eq (1):
$$ g(s)= -i \langle \pi ^{0} K^{-}
\vert \partial ^{\mu} V_{\mu}^{4-i5}(0) \vert 0 \rangle =
(m_K^2-m_\pi^2)f_1(s)+sf_2(s)\kern 1.5 cm (2) $$ g(s) is therefore a pure
scalar
which describes the S wave $\pi K$ form factor.  g(s) measures the $SU(3)$
violating effect because, in the exact SU(3) limit, the vector current is
conserved. We expect therefore in the  $\tau
\to \pi K \nu$ decay, the P wave form factor $f_1(s)$ dominates.

Because of the octet current hypothesis the two channels $\pi^0 K^-$ and $\pi^-
\bar{ K^0}$ matrix elements are related by the  Clebsh Gordon coefficient
$$\langle\pi^- \bar K^0\vert V_{\mu} ^{4-i5}(0)\vert0\rangle=\sqrt{2}
\langle\pi^0  K^-\vert V_{\mu} ^{4-i5}(0)\vert0\rangle\kern 1.5 cm (3) $$
In terms of form factors $f_1(s)$ and $g(s)$, and the angle  $\theta$, defined
as angle between $ \vec p_\pi$  and $ \vec p_\nu$  in the hadronic rest frame,
 the decay spectrum and forward backward asymmetry are given by:

$$\eqalign{ {d \Gamma\over dsdcos\theta}
&={  G_F^2\sin^2 \theta_c(m_{\tau}^2-s)^2\lambda^{1/2}(s,m_{\pi}^2,m_k^2)
 \over 2^9\pi^3 s m_{\tau}^3 } \{ {\lambda/ s} {\vert f_1(s)
\vert} ^2\sin^2 \theta \cr
&\qquad\qquad
+{ m_\tau^{2} \over s^2} \vert {g_0(s)
}+{\lambda^{1/2} f_1(s)\cos\theta  } \vert ^2
\bigr \} }\eqno (4)$$
where $\theta _{c}$ is the Cabibbo angle  with $\cos {\theta_c}=0.97 $, and $
\lambda(s,m_\pi^2,m_k^2) = (s-(m_\pi+m_K)^2) (s-(m_\pi-m_K)^2) $. For
 simplicity we denote it by  $\lambda $.

The forward-backward asymmetry is defined as:

$$ \eqalign{ A_{FB} & ={{ d\Gamma[\cos \theta]-d\Gamma[-\cos \theta]}\over
{d\Gamma[\cos\theta]+ d\Gamma[-\cos\theta]}} \cr
& =
{\lambda^{1/2} m_\tau^2 Re[g*(s) f_1(s)] \over s [\lambda (2/3+{m_\tau^2 /
3 s}) {\vert f_1(s)\vert} ^2 +{m_\tau^2}{\vert g(s) \vert}^2/s]} }\eqno (5)$$

The forward-backward asymmetry is a useful phenomenological quantity, because
it
allows us to measure experimentally the relative phase  between S and P
wave amplitudes of the $\pi K$ scattering.  Furthermore, because the
forward-backward
asymmetry vanishes in the limit of the exact SU(3) symmetry, the presence of
this term allows us to measure the SU(3) breaking effect.

For completeness, we now
add a short review of the $K_{l3}$ decay. The total amplitude is a product
of two parts, the
leptonic  and the hadronic ones. The hadronic matrix element is given by:

$$\eqalign {&\langle \pi ^{0}( p_1) \vert V_{\mu}^{4+i5}(0) \vert
K^{-}(p_2) \rangle  =f_+(t)(p_1+p_2)_\mu +
f_-(t) ( p_2 - p_1)_\mu\cr
&f_0(t) = i \langle \pi \vert \partial ^{\mu} V_{\mu}^{4+i5}(0) \vert K
\rangle =
(m_K^2-m_\pi^2)f_+(t)+t f_-(t) }\eqno (6)$$
$f_+$ and $ f_- $ are dimensionless form factors depending on the
momentum transfer $t=( p_2 -p_1)^2$, they are respectively  the analytic
continuation of
$f_1$ and $f_2 $. $K_{\mu 3}$ experiments  give information on $f_+$ and
$f_-$, while  $K_{e
3}$ experiments are sensitive only to $f_+$ because of the small electron mass.

Using the Ademollo-Gatto theorem $f_+(0)={1/ \sqrt 2}$ and hence\hfill\break
 $f_0(0) ={ (
m_K^2-m_\pi^2)/ \sqrt 2}$ for $\pi^0 K^-$ system.

Using the standard  current algebra technique and the $SU(2)_L \times SU(2)_R $
commutation relation by taking the pion momentum $p_1$ soft we have the
well known
 Callan-Treiman relation [\callan]:
$$ f_+(m_K^2) +f_-(m_K^2) ={ f_K \over f_\pi \sqrt 2}\kern 1.5 cm (7)$$
where $f_K$ and  $f_\pi$ are, respectively, the K and $\pi$ decay constants
${f_K/f_\pi}=1.22$.

By evaluating Eq(2) at $t=m_K^2$ and noting that $f_-(m_K^2)$ is proportional
to ${ m_{\pi}^2/ m_{K}^2 }$ we
have: $f_0(m_K^2) \approx   f_0(0){ f_K/f_\pi}$

\vskip 1.0 cm
 \centerline {\bf {II) Unitarity  correction to current algebra
 results}}
\vskip 0.5 cm

Before giving the details of the one loop calculation, we  outline some
general properties of the S matrix.

  The elastic unitarity condition, which should be valid in the physical
region of the
$\tau\to\pi K\nu$ decay gives:
$$\eqalign { Imf_1(s)&=f_1(s)\exp {-i\delta_p^{1/2} }\sin {\delta_p^{1/2}} \cr
 Img(s)&=g(s)\exp {-i\delta_s^{1/2} }\sin {\delta_s^{1/2}}    }\kern 1.5 cm
(8)  $$
where $\delta_s^{1/2}$ and $\delta_p^{1/2}$ are respectively the phase of S and
P wave I=1/2 $\pi K$ scattering amplitude. One can decompose $\pi K$
elastic amplitude into the partial
waves using
$T^{I}(s,\theta) =16\pi\sum_l (2l+1)t_l^I(s)P_l(\cos \theta)$, where
l stands for the angular momentum and $\theta$ the angle in c.m system,
and $t_l^I(s)={\exp {i\delta_l^{I}
}\sin {\delta_l^{I}}/ \rho (s)}$ where $\rho (s)={ \sqrt {\lambda (s,m_\pi
^2,m_K^2)}/ s}$ is the phase space  factor. From Eq(8) in order to satisfy
the elastic
unitarity condition the form factors $f_1(s)$ and g(s) must have,
respectively, the
phase $\delta_p^{1/2}$ and $\delta_s^{1/2}$.

The elastic unitarity condition is a good approximation to describe $\tau
\to \pi K \nu$
 decay
owing to the experimental fact that the inelastic effects are not large.
 Using the analytic properties of the form factors, and the elastic
unitarity condition, we have:

$$\eqalign { & f_1(s) =f_1(0)+f_1(0){\langle r_v^2
\rangle\over 6}s+{s^2\over\pi}\hskip -5mm\int\limits_{(m_\pi+m_K)^2}
^{+\infty}\hskip -4mm{f_1(z)\exp {-i\delta_p^{1/2} }\sin
{\delta_p^{1/2}}dz\over
z^2(z-s-i\epsilon)} \cr
&
g (s) =g(0)+g(0){\langle r_s^2
\rangle\over 6}s+{s^2\over\pi}\hskip -5mm\int\limits_{(m_\pi+m_K)^2}
^{+\infty}\hskip -4mm{g(z)\exp {-i\delta_s^{1/2} }\sin {\delta_s^{1/2}}dz\over
z^2(z-s-i\epsilon)} } \eqno (9) $$
where $\langle r_v^2\rangle$ and $\langle r_s^2\rangle$ are, respectively,
vector and scalar radii of $\pi K $ system.

The solutions to these integral equations are well known [\omnes]
$$ \eqalign{  &f_1(s) = f_1(0)\exp ({s\over \pi}\hskip
-4mm\int\limits_{(m_\pi+m_K)^2
} ^{+\infty}\hskip -3mm {
\delta_p^{1/2}dz\over z(z-s-i\epsilon})\cr
&
g(s) = g(0)\exp ({{s\over \pi}\hskip -4mm\int\limits_{(m_\pi+m_K)^2}
^{+\infty}\hskip -3mm {
\delta_s^{1/2}dz\over z(z-s-i\epsilon}}) }\eqno (10) $$
These solutions can also be derived by an infinite iteration of the
integral equations
Eq[9].

 From the Ademollo-Gatto theorem as discussed above,  we have $f_1(0)= {1/
\sqrt 2}$ and hence $g(0) ={ ( m_K^2-m_\pi^2)/ \sqrt 2}$.
We ignore the so called polynomial ambiguity which is obtained by multiplying
the rhs of Eq [10] by a polynomial. They correspond to higher energy
contributions
 which are assumed to be small.
We can either use the experimental or theoretical phase shifts in Eq [10]
to calculate the form
factors $f_1(s)$ and g(s).

A simplest approximation for $f_1(s)$ and g(s) can be obtained by modifying
the one loop CPTh as was
done in the reference [\tnts] for the pion form factor. This can be done by
calculating
the strong $\pi K$ $I=1/2$ amplitude. The tree amplitudes are:

$$\eqalign{& t_1^{tree}(s) ={ \lambda (s,m_\pi ^2,m_K^2) \over 128\pi
sf_\pi ^2}\cr
&
t_0^{tree}(s)= {(2s-{3\lambda (s,m_\pi ^2,m_K^2)/(
4s)}-2m_\pi^2-2m_K^2)\over 32\pi f_\pi^2} }\eqno (11)$$
where the subscript refers to the l partial wave.

Using these expressions and replacing $f_1(z)$ by $f_1(0)$ and g(z) by g(0)
in Eq[9], we have the one
loop perturbative results for the form factor $f_1$ and g:
\vskip 0.5 cm

$$\eqalign{ &f_1^{pert.}(s) = f_1(0)+f_1(0) {\langle r_v^2\rangle\over
6}s+{f_1(0)\over128\pi f_\pi^2}(-I_1(s)+2(m_\pi^2+m_K^2)I_2(s)-\cr
&
\hskip 1.9cm (m_\pi^2-m_K^2)^2I_3(s)) }\eqno (12-a)$$
$$\eqalign{& g^{pert.}(s) = g(0)+g(0) {\langle r_s^2\rangle\over 6}s
+{g(0)\over 32\pi
f_\pi^2}(-{5\over 4}I_1(s)+ {1\over
2}(m_\pi^2+m_K^2)I_2(s)+\cr
&\hskip 1.9cm{3\over4}(m_\pi^2-m_K^2)^2 I_3(s)) }\eqno (12-b)$$
where $I_1(s)$, $I_2(s)$ and $I_3(s)$ are given in the Appendix A.
As was explained in reference [\tnts], these expressions only satisfy
perturbatively the unitarity
relation. We can resum the perturbative results, Eq[12-a, 12-b], to
implement the elastic unitarity relation.
For this purpose, following ref[\tnts], we write $ f_1$ and g as:

$$f={f^{tree}\over{ 1-{f^{loop}/f^{tree}}}}\eqno (13)$$
which is just the diagonal [1,1] Pad\'{e} approximant of the form factors,
hence:

$$f_1(s)= {f_1(0)\over1-{s \langle r_v^2\rangle/
6}-{1\over128\pi f_\pi^2}(-I_1(s)+2(m_\pi^2+m_K^2)I_2(s)-(m_\pi^2-m_K^2)^2
I_3(s))}\eqno (13-a)$$
$$g(s)={g(0)\over 1-{s \langle r_s^2\rangle/6} -{1\over 32\pi
f_\pi^2}(-{5\over 4}I_1(s)+ {1\over
2}(m_\pi^2+m_K^2)I_2(s)+{3\over4}(m_\pi^2-m_K^2)^2 I_3(s))}\eqno (13-b)$$

We show below that these results can also be directly obtained  by the
$N/D$ method in the approximation
 where the t and u one loop graphs are represented by an adjustable
polynomial in the D function
[\bel] or by the once iterated solution of the integral equation for the
inverse of the form
factor. It is  obvious that Eq(13) can also be obtained by the infinite
bubble summation of the
$\pi K$ S and P wave interactions.

Because the partial wave amplitude has both right and  left hand cut, we
can always write it as
a product of two cuts;
$ t_l^{I}(s) ={N_{l}^{I}(s)/D_{l}^{I}(s)}$,
  we normalize $D_{l}^{I}$ such that $D_{l}^{I}(0)=1$.
The elastic unitarity implies $Im(t_l^{I}(s)) = \rho(s)  \vert t_l^{I}(s)
\vert ^2 $
and hence  $Im(D_{l}^{I}(s)) = -\rho(s)N_{l}^{I}(s)$.
Using analyticity and unitarity,  we can write the
following dispersion relation for the partial wave amplitude  $$
t_l^{I}(s)={N_{l}^{I}(s)\over 1+s D^{\prime}(0)-{s^2\over \pi}
\int\limits_{(m_\pi+m_K)^2} ^{+\infty}{\rho(z)N_{l}^{I}(z) dz\over
z^2(z-s-i\epsilon)}}\eqno (14)$$
 $D^{\prime}(0)$ is an adjustable phenomenological parameter. We shall
approximate
 $N_{l}^{I}(s)$ by $t_l^{tree}(s)$. Because $1/ D_{l}^{I}(s)$  has the
following
phase representation: $1/ D_{l}^{I}(s)= \exp ({s\over \pi}\hskip
-3mm\int\limits_{
(m_\pi+m_K)^2} ^{+\infty}\hskip -2mm {
\delta_{l}^{I}dz\over z(z-s-i\epsilon})$, hence $1/ D_{1}^{1/2}$ and $1/
D_{0}^{1/2}$  are proportional to $f_1(s)$ and
g(s) given by Eq[10], and hence we have:
$$  f(s)={f(0)\over 1+s D^{\prime}(0)-{s^2\over \pi}
\int \limits_{(m_\pi+m_K)^2} ^{+\infty}{\rho(z)t_l^{tree}(z) dz\over
z^2(z-s-i\epsilon)}}\eqno (15)$$

This expression is  equivalent to (13-a) and (13-b) if we identify $
D^{\prime}(0)$ with the rms radius.

 From the expression for $f_1(s)$, the phase of the form factor which is
identical to the
P wave phase shifts of $\pi K$ scattering amplitude, can be calculated
using the experimental value of $ \langle r_v^2\rangle\
=0.34 \pm 0.03 fm^2$. Using this value we have $m_{K^*}=810\pm 30$ MeV and
agrees with
the experimental data $m_{K^*}=892$MeV. Its width satisfies the following
modified
KSRF relation [\ksrf]:$$\Gamma _{K^*}={
\lambda^{3/2}(m_{K^*}^2,m_\pi ^2,m_K^2) \over 128 \pi m_{K^*}^3
f\pi^2}\eqno (16)$$
Using the experimental value $m_{K^*}=892$MeV the numerical result of the
right hand side of Eq(16) is $55$ MeV,
compared to the experimental value of $49.8 \pm 0.8$ MeV.

The branching ratio $B.R = {\Gamma( \tau \to \pi K \nu ) \over \Gamma(\tau
\to all)}$ is $ 1.0 \% $ and is in agreement with the experimental result
of $ B.R_{exp.}= (1.4 \pm 0.2) \% $

Because the S wave $\pi K$ scattering length does not vanish, g(s) has a
square root threshold
singularity at the threshold (the derivative of g(s) is discontinuous at
this point) as it
can be seen in Fig(3).
The scalar form factor contributes very little to the $\pi K$ spectrum
owing to the fact that it appears as a square of the amplitude.
The forward backward asymmetry, being proportionnal to the amplitude, is
reasonably large. It is about
$10\%$ in the $K^{*}$ resonance region where the number of event is
maximum. The
Forward-Backward asymmetry could be a useful quantity
for studying the relative phases of the S and P waves as can be seen from
Eq(5).

\vskip 0.5 cm
\centerline { {\bf III) A more exact calculation of $\pi K \to \pi K$
scattering and
form factors }}
The $\pi K$ scattering problem was calculated up to the one loop order by
ref[\meisner]
and will not be repeated here.
 Below the inelastic thresholds, the one loop chiral perturbation theory
satisfies
the perturbative unitarity: Im $t^{(1)}=\rho (s) {t^{(0)}}^2 $ where $
t=t^{(0)}+
t^{(1)} $ and the superscripts stand for the tree graph and one loop
calculation, the isospin and the partial wave indices are omitted for
convenience.
This relation is not really
satisfied in the one loop amplitude of ref[\meisner]
unless we replace $f_{K}$ by $f_{\pi}$. In fact, At the order  $0(p^4)$,
it is not clear
whether  $f_{\pi}$ or $f_K$
should be used. The difference is of the order $O(p^6)$. The way to
circumvent this problem is to
calculate the $\pi K$
scattering in the SU(2) $\times$ SU(2) theory which  is not yet available.
Using the standard
current algebra technique [\weinbf] by treating the K meson as a heavy
target one expect
the $\pi K\to\pi K$ amplitude to be proportionnal to $1/{f_{\pi}^2}$. In
what  follows
we replace $f_{K}$ by $f_{\pi}$ in order to get the same absorptive part in
the s channel
 as in our
calculation in section 2.  It is staightforward
to show that the reconstructed amplitude:

$$ t(s)={t^{(0)}\over 1-{t^{(1)}/t^{(0)} } } \eqno (17)$$ satisfies exactly
the elastic
unitarity. The counterterms for the one loop amplitude are discussed in
ref[\gasser].
 $L_4$,
$L_5$, $L_6$ and $L_8$ measure the chiral symmetry breaking effects; their
contributions to the
scattering amplitude are proportional to the pion and Kaon mass squared.
In this work they are taken to be the
 values given by the reference [\gasser]. One linear combination of
$L_{1}, L_2$ and $L_3$ is given by the
$K^{*}$ mass. The other two  constraints were considered in ref[\bel] in
studying   $
\pi\pi$ scattering: $L_{2}-2L_{1}-L_3$ is fixed
by the $\rho$ mass, the third constraint, in order to completely
determine $L_{1}, L_2$ and $L_3$,
was obtained from an experimental S wave I=0 $\pi\pi$ phase shift at
500 MeV. The $K^{*}$ and $\rho$ mass are defined, respectively, as the
energies where I=1/2 l=1 $\pi$
K and I=1 l=1 $\pi\pi$ partial wave phase shifts passe through $90$
degrees. These three constraints
give $L_1=1.23\times10^{-3}$, $L_2=1.51\times 10^{-3}$ and
$L_3=-4.1\times10^{-3}$.

 The predictions
of the scattering lengths are: $a_0^{1/2}=0.22 m_{\pi}^{-1}$, and
$a_1^{1/2}=0.016 m_{\pi}^{-3}$.
They are
in agreement with experiments. The  P wave phase I=1/2 phase shifts are given
in
Fig(2). It is seen that they are in a reasonable agreement with the
experimental data
[\estab] and also with
the results calculated above by N/D method where the the left hand cut
discontinuity is neglected.
More explicitly for the same value of the $K^{*}$ mass, this more complete
calculation yields $\Gamma _{K^{*}
}=45 MeV$, while the N/D method gives $\Gamma _{K^{*}}=55 MeV$.  If we had
taken the $\pi K\to
\pi K$ one loop amplitude to be inversely proportionnal to
$1/{f_{K}^{2}f_{\pi}^{2}}$ with
$f_{K}=1.22 f_{\pi}$,
we would have obtained $\Gamma _{K^{*}}=38 MeV$ which is too small. The
corresponding form factors using the phase
representation are shown in Fig(1). It is seen that the peak value of the
vector form factor
squared in the present calculation, where the left hand cut in the $\pi K$
scattering
 amplitude is taken into account, is 25\% higher than that
obtained from N/D method where the left hand cut is neglected.  The
discrepancy is due
to the difference in $ K^{*}$ width  obtained in these two
calculations. Away from the peak, the approximate solution agrees well with
the more
exact calculation.

This more exact calculation yields a branching ratio of $1.15\%$ for
$\tau \to\pi K\nu$ decay which is in a better agreement with the
experimental value $1.4\pm 0.2 \%$.

The S wave phase shift calculated using the unitarized $\pi K$ amplitude
eq(17), agrees
also better with the
experimental data [\estab, \exps].  The approximate S wave phase shift
eq(13-b) where the left
hand cut of $\pi K$ is neglected, differs from the
experimental results at high energies.

Our result shows that the calculated S and P wave form factors, using the
usual rule
of neglecting  the left
hand cut for the $\pi K$ scattering are not always accurate. When
the left hand cut of the $\pi K$ scattering is taken into account, a better
agreement with the experimental data is obtained. In other words, there are
some sizable
corrections to the large $N_f$ expansion.

We can also calculate the vector
and scalar $\pi K$ rms radii using:

$$ \eqalign{  &{\langle r^2_{V}\rangle}^{\pi K}={6\over \pi}
\int\limits_{(m_\pi+m_K)^2} ^{+\infty} {
\delta_1^{1/2}dz\over z^2}\cr
&{\langle r^2_{S}\rangle}^{\pi K}={6\over \pi} \int\limits_{(m_\pi+m_K)^2}
^{+\infty} {
\delta_0^{1/2}dz\over z^2 } }\eqno (18) $$

Numerical calculation using the $\pi K$ phase shift from the exact calculation
gives $\langle r^2_{V}\rangle ^{\pi K}=0.27 fm^2$ and
$\langle r^2_{S}\rangle ^{\pi K}=0.13 fm^2 $. The experimental data from
$K_{\mu 3}$
decay are  $\langle r^2_{V}\rangle ^{\pi K}=0.34\pm 0.03 fm^2$. The
experimental
 situation of
 the $\langle r^2_{S}\rangle ^{\pi K}$[\pdg] is unsatisfactory since the
values obtained
by the expriments[\expf,\donald] are quite dispersed. The best result is given
by Donaldson et al.
[\donald], $\langle r^2_{S}\rangle ^{\pi K}=0.23 \pm 0.05 fm^2 $ which is
larger than our
theoretical prediction $\langle r^2_{S}\rangle ^{\pi K}=0.13 fm^2 $. The
reason for this
discrepancy is due to the assumption of the linear dependence in s of the
form factor
in the analysis of the experimental data of $K_{\mu 3}$. Our scalar form
factor calculation
disagrees with this assumption as can be seen from Fig(3).

In this article we have calculated the $\pi K$ S and P wave form factors by
two different
methods. In the first method, using the input as the $\pi K$ r.m.s radius
and the
bubble summation for the form factor, the calculated  form factor moduli
and phases are in a
rough agreement with the experimental data.

In the second method, we calculate first the CPTh for the $\pi K$
scattering and then we
unitarize this amplitude (where both left and right cuts are
included) by the Pad\'e Approximant method;  the strong $\pi K$ elastic
amplitudes
are in
good agreement with the experimental data (e.g $\pi K$ phase shifts, width
and mass
of the $K^{*}$ resonance).
We then calculate the $\pi K$
S and P wave form factors using the Omne\`s representation. This method yields
 a better agreement with data than the first one.

 \vskip 1.0 cm

\centerline {\bf {APPENDIX A}}
\vskip 0.5 cm
In section 2 we have given the form factors in terms of $I_1(s), I_2(s)$
and $I_3(s)$ which are easily expressed in terms of a generating function;
$$ \psi(s)=-{\lambda(s,m_\pi^2,m_K^2)\over
2}\int\limits_{(m_\pi+m_K)^2}^{\infty}{dz \over \sqrt
{\lambda(z,m_\pi^2,m_K^2)}(z-s-i\epsilon)}$$

For convenience we give the analytic continuation of this function to all
regions.

$$ \psi(s) =\cases { $$\eqalign { &\sqrt {\lambda(s,m_\pi^2,m_K^2)}\log({{\sqrt
{s-(m_\pi+m_K)^2}+\sqrt { s-(m_\pi-m_K)^2}}\over 2\sqrt {m_\pi m_K}})\cr
 &
 -i{\pi \over 2}\sqrt {\lambda(s,m_\pi^2,m_K^2)} } $$&if$s\ge s_{t}$\cr
-\sqrt {\lambda(s,m_\pi^2,m_K^2)}\log({{\sqrt
{-s+(m_\pi+m_K)^2}+\sqrt { -s+(m_\pi-m_K)^2}}\over 2\sqrt {m_\pi m_K}})&if$
s\le s_{t}$
\cr
\sqrt {\vert \lambda(s,m_\pi^2,m_K^2)\vert}\arctan(\sqrt{ {
s-(m_\pi-m_K)^2}\over{-s+(m_\pi+m_K)^2}})&if not\cr }  $$
where $s_{t}=(m_\pi+m_K)^2$

$$ I_1(s) = -{ s^2\over \pi}\int \limits_{(m_\pi+m_K)^2}^{\infty} {\sqrt
{\lambda(z,m_\pi^2,m_K^2)} \over z^2(z-s-i\epsilon)}dz $$
$$ I_2(s) = -{ s^2\over \pi}\int \limits_{(m_\pi+m_K)^2}^{\infty} {\sqrt
{\lambda(z,m_\pi^2,m_K^2)} \over z^3(z-s-i\epsilon)}dz $$
$$ I_3(s) = -{ s^2\over \pi}\int \limits_{(m_\pi+m_K)^2}^{\infty} {\sqrt
{\lambda(z,m_\pi^2,m_K^2)} \over z^4(z-s-i\epsilon)}dz $$

$$I_1(s)={2\over \pi}(\psi(s)-\psi(0)-s\psi^\prime (0))$$
$$I_2(s)={2\over \pi s} (\psi(s)-\psi(0)-s\psi^\prime (0) -{s^2\over
2}\psi^{\prime\prime}(0))$$ and $$I_3(s) ={2\over \pi s^2}
(\psi(s)-\psi(0)-s\psi^\prime (0) -{s^2\over 2}\psi^{\prime\prime}(0)-{s^3\over
6}\psi^{\prime\prime\prime}(0))$$

\centerline{\bf {REFERENCES}}

\vskip 0.5 cm
\item{[{\adler}]} S.L Adler {\it Phys.Rev.}140,{\bf B736} (1965)\hfill\break
W.I Weisberger{\it Phys.Rev.}143,1302 (1966).\hfill\break
\item{[{\weinbf}]} S.Weinberg,{\it Phys.Rev.Lett.},17,616(1966).\hfill\break

\item{[{\weinbs}]} S.Weinberg, {\it Phys.Rev.Lett.},17,336(1966).\hfill\break
\item{[{\addashen}]} For a complete list of references see S.L Adler and
R.F Dashen, {\it Current Algebras and applications to particle physics},
W.A.Benjamin, Inc. (1968). \hfill\break
\item{[{\schnitzer}]} S. Gasiorowicz and D.A. Geffen, {\it Rev.Mod.Phy.},
{\bf 41}, 531(1969). \hfill\break
I.S Gerstein, H.J Schnitzer and S.Weinberg,
{\it Phy.Rev.} {\bf 5},1873(1968).\hfill\break
I.S Gerstein, H.J Schnitzer,
{\it Phy.Rev.} {\bf 5},1876(1968).\hfill\break

\item{[{\bels}]} L. Beldjoudi and T.N. Truong, Ecole Polytechnique preprint
CPTH-A336.1194   to be published in {\it Phy. Lett.} {\bf B}.
\hfill\break
\item{[{\tntf}]} T.N. Truong, {\it Phy. Lett.} {\bf B}99(1981)154.\hfill\break

\item{[{\im}]}C.J. Im, {\it Phy. Lett.} {\bf B}281(1992)357.\hfill\break
A.Dobado and J.R Pelaez, {\it Phy. Lett.} {\bf B}286(1992)136.\hfill\break
T.N. Truong, {\it Phy. Lett.} {\bf B}313(1993)221.\hfill\break

\item{[{\tnts}]}  T.N. Truong,{\it Phy.Rev.Lett.} {\bf
61},2526(1988).\hfill\break

\item{[{\callan}]} C.Callan and S.Treiman,{\it
Phys.Rev.Lett.},16,153(1966).\hfill\break
V.S Mathur, S.Okubo and L.Pandit, {\it Phy.Rev.Lett.} {\bf 16},371(1966),
601(E).
\hfill\break
M.Suzuki, {\it Phy.Rev.Lett.} {\bf 16},212(1966).
\hfill\break

\item{[{\bel}]} L. Beldjoudi and T.N. Truong, Ecole Polytechnique preprint
CPTH-A292.0294.
\hfill\break

\item{[{\omnes}]}  N.I.Muskhelishvili,{\it Singular Integral
Equations(Noordhoof,Groningen,1953)}

       R.Omnes,{\it Nuovo Cimento} 8,316(1958).\hfill\break

\item{[{\ksrf}]} K.Kawarabayashi and M. Suzuki, { \it Phy. Rev. Lett.} {\bf
16}(1966)255.\hfill\break
Riazuddin and Fayyazudddin,  {\it Phy. Rev.}{\bf 147}(1966)1071.
\hfill\break

\item{[{\meisner}]} V.Bernard, N. Kaiser and U.G Mei$\beta$ner
{\it Nucl. Phy.} {\bf B}357(1991)129.
\hfill\break

\item{[{\marshak}]} For more details see R.E. Marshak, Riazuddin and C.P.
Ryan, {\it Theory of Weak Interactions
in
Particle Physics}, Wiley-Interscience (1969). \hfill\break

\item{[{\gasser}]} J.Gasser and H. Leutwyler, {\it Ann.Phy. (NY)}158(1984)
142;\hfill\break
{\it Nucl. Phy.} {\bf B}250 (1985)465.\hfill\break

\item{[{\estab}]} P. Estabrooks et al., {\it Nucl.Phy. }B133(1978)
490.\hfill\break

\item{[{\exps}]} M.J. Matison et al., {\it Phy. Rev.} {\bf D9}(1974)1872.
\hfill\break
                 H.H. Bingham et al., Nucl.Phy.B41 (1972)1.\hfill\break
                 R. Mercer et al., Nucl. Phy. B32(1972)381.\hfill\break

\item{[{\pdg}]} Particle Data Group, Phy. Lett. B111(1982).\hfill\break
\item{[{\expf}]} V.K Birulev et al. {\it Nucl. Phy.} {\bf B}182(1981)1.
\hfill\break
Y.Cho et al., {\it Phy. Rev.} {\bf D22}(1980)2688.  \hfill\break
D.G Hill et al., {\it Nucl. Phy.} {\bf B}153(1979)39.
\hfill\break
A.R Clark et al., {\it Phy. Rev.} {\bf D15}(1977)553.  \hfill\break
C.D Buchanan et al., {\it Phy. Rev.} {\bf D11}(1975)457.  \hfill\break

\item{[{\donald}]} G. Donaldson et al., {\it Phy. Rev.} {\bf D9}(1974)2960.
 \hfill\break

\vskip 4.0 cm
\centerline{{\bf FIGURE CAPTIONS}}

\item{{\bf Figure 1}} : The calculated $\pi K$ P wave form factors
(solid/dashed/dot-dashed curves) corresponding
 respectively to, the Omn\'es representation using the $\pi K$ phase shift
calculated from the
unitarized
one loop CPTh as defined in Eq(10), the [1,1] Pad\'e approximant as given
in Eq(13-a),
CPTh prediction as given in Eq(12-a).\hfill\break
\item{{\bf Figure 2}} :

The solid line represents the I=1/2, l=1 $\pi K$ scattering phase shift
calculated from
the unitarized CPTh Eq(17).
The dashed line corresponds to the similar phase shift when the left
 hand cut is neglected as given by eq(13-a).
 The dot-dashed line is the CPTh prediction phase Eq(12-a)(which is not the
same
as the phase shift due to the violation of the full elastic unitarity relation
in this
method). The experimental results are those of ref.
[\estab].\hfill\break
\item{{\bf Figure 3}} :The calculated $\pi K$ S wave form factors
(solid/dashed/dot-dashed curves) corresponding
 respectively to, the Omn\'es representation using the $\pi K$ phase shift
calculated from
 unitarized one
loop CPTh as defined in Eq(10), the [1,1] Pad\'e approximant
as given in eq(13-b), CPTh prediction as given by Eq(12-b).\hfill\break

\item{{\bf Figure 4}}:
The solid line represents the I=1/2, l=0 $\pi K$ scattering phase shift
calculated from
the unitarized CPTh Eq(17).
The dashed line corresponds to the similar phase shift when the left hand
cut is
neglected as given by eq(13-b).
 The dot-dashed line is the CPTh prediction phase Eq(12-b)(which is not the
same
as the phase shift due to the violation of the full elastic unitarity
relation in this
method). The experimental results are those of ref.
[\estab, \exps].\hfill\break
\item{{\bf Figure 5}}:
Calculation of the $\pi K$ invariant mass squared spectrum of
$\tau \to \pi K \nu$ decay. The dashed/solid curves correspond respectively
to the calculations
with/without the left hand cut of $\pi K$ scattering amplitude. \hfill\break

\item{{\bf Figure 6}} :

Prediction for the Forward-Backward asymmetry $A_{FB}$ defined in eq(5) as
a function of the $\pi K$ invariant mass squared. The dashed/solid curves
correspond respectively to the calculations
with/without the left hand cut of $\pi K$ scattering amplitude.\hfill\break

\vskip 6 cm
\end